# Resonant excitation of quantum emitters in hexagonal boron nitride

*Toan Trong Tran,[1,*] Mehran Kianinia,[1] Minh Nguyen,[1] Sejeong Kim,[1] Zai-Quan Xu,[1] Alexander Kubanek,[2,3] Milos Toth,[1,*] and Igor Aharonovich[1,*]*

[1]School of Mathematical and Physical Sciences, University of Technology Sydney, Ultimo, NSW, 2007, Australia

[2]Institute for Quantum Optics, Ulm University, Albert-Einstein-Allee 11, D-89081 Ulm, Germany

[3]Center for Integrated Quantum Science and Technology (IQST), Ulm University, Albert-Einstein-Allee 11, D-89081 Ulm, Germany

ABSTRACT

Quantum emitters in layered hexagonal boron nitride (hBN) have recently attracted a great attention as promising single photon sources. In this work, we demonstrate resonant excitation of a single defect center in hBN, one of the most important prerequisites for employment of optical sources in quantum information application. We observe spectral linewidths of hBN emitter narrower than 1 GHz while the emitter experiences spectral diffusion. Temporal photoluminescence measurements reveals an average spectral diffusion time of around 100 ms. On-resonance photon antibunching measurement is also realized. Our results shed light on the potential use of quantum emitters from hBN in nanophotonics and quantum information.

TEXT

Solid-state quantum emitters in low-dimensional hosts have emerged as promising candidates for quantum information and communications, owing to their strong photoluminescence (PL) and the potential use in integrated nanophotonics.[1-4] Within this class of single photon emitters, optically active defect centers in layered hexagonal boron nitride (hBN), a two-dimensional dielectric, have gained tremendous research momentum due to outstanding characteristics such as superb brightness at room temperature,[5-7] high photon contribution into zero phonon lines (ZPL),[8,9] linearly polarized emission,[10-12] high photo-stability even upon heating to 800 K,[13] and spectral tunability.[7,8] Furthermore, integration of hBN emitters with plasmonic nanocavities[14] and tapered-fibers[15] have recently been demonstrated as a first step towards integrated on-chip circuits.

Recent cryogenic measurements revealed that some of the emitters have stable spectral lines as narrow as 45 μeV.[9,16] It has also been shown that some emitters suffer from ultrafast spectral diffusion that causes broadening of the ZPLs.[17] However, to date, coherent resonant excitation of these quantum emitters has not been demonstrated. Resonant excitation is important for understanding the fundamental photophysical processes of solid state quantum emitters, and vital for realization of advanced quantum experiments, including generation of indistinguishable photons, entanglement and optical coherent control of quantum states.[4,18-24]

In this work, we report resonant photoluminescence excitation (PLE) of a single hBN emitter at 8 K. The emitter shows optical linewidths of less than 1 GHz, but blinking and spectral diffusion result in a broader optical envelope that spans approximately 6.3 GHz. Despite the spectral diffusion, high purity single photon emission is confirmed by recording an on-resonance second-order autocorrelation function. Our results shed light on outstanding challenges with this intriguing quantum system, and represent a stepping stone towards the generation of indistinguishable photons for quantum information processing applications.

We employed hBN flakes (Graphene Supermarket, ~200 – 500 nm in diameter) that were drop-casted on a 1 x 1 cm$^2$ silicon substrate. The substrate was thermally treated in argon at 850°C for half an hour to optically activate defects in hBN.[5] The sample was mounted on a three-dimensional (3D) piezo stage (Attocube Inc.) of a home-built open-loop cryostat with flowing liquid helium (figure 1a). The sample was excited with a computer-controlled continuous-wave (CW) Titanium:Sapphire (Ti:Sap) laser (SolsTis, M2 Inc.) with a narrow spectral linewidth of 50 kHz. Excitation and collection light were split by a 90:10 (transmission: reflection) non-polarizing beamsplitter and collected through an objective lens with (NA= 0.95), which was mounted inside the cryostat. Residual pump laser was rejected using the combination of a 715 nm longpass and 850 ± 105 nm bandpass filter (Semrock Inc.). It is noted that for both PLE and on-resonance excitation experiments, only the light from the phonon sideband (PSB) was collected, and not from the ZPL, to avoid collection of scattered laser light. Time-resolved PL was carried out with a pulsed 675 nm diode laser (~ 50 ps pulse width) with a repetition rate of 10 MHz and a power of 100 µW. Second-order autocorrelation measurements were performed with a Hanbury Brown and Twiss (HBT) interferometer and a time-correlated single photon counting (TCSPC) module.

Quantum emitters in hBN flakes are known to display a wide range of ZPL energies spanning ~ 570 – 770 nm,[8, 10, 11, 25, 26] making it possible to select and address a particular optical transition of interest within this range. We therefore conducted survey confocal PL mapping using a CW laser tuned at 700 nm to off-resonantly excite hBN emitters at 8 K, and selected a bright emitter for further optical investigation (figure 1b, black arrow). The emitter has a weak phonon sideband (PSB) with an estimated Debye-Waller (DW) factor of $DW = \frac{I_{ZPL}}{I_{tot}} = 0.9 \pm 0.1$, and an asymmetric ZPL shape, consistent with the literature.[8] While the full-width-at-half-maximum (FWHM) of the emitter at room temperature was ~10 nm (5.1 THz), at 8 K, the ZPL width was limited by our spectrometer resolution of 25 GHz (figure 1c). It should be noted that the small peak at ~ 800 nm is a ZPL of another weak emitter within the excitation spot. The inset of figure 1c shows a higher resolution spectrum of the emitter, indicating a ZPL position of 766.8 nm (391.2 THz).

To verify the single photon purity of the defect under off-resonant excitation, we recorded the second-order autocorrelation function, $g^{(2)}(\tau)$. First, to quantify the jitter contribution in our detection scheme, we measured the standard deviation of the instrument response function (IRF) to be 0.5 ns by fitting a Gaussian on the IRF. We then used the Gaussian-convoluted three-level model to fit the data, taking into account the timing-jitter effect:[27]

$$g^{(2)}_{meas}(\tau) = \int_{-\infty}^{+\infty} g^{(2)}(\tau') J(\tau - \tau') d\tau' \quad (1)$$

with $g^{(2)}(\tau') = 1 - (1+a)e^{-\tau'/\tau_1} + ae^{-\tau'/\tau_2}$ **(2)**, and $J(\tau - \tau') = \frac{1}{\sigma\sqrt{2\pi}} e^{\frac{-(\tau-\tau')^2}{2\sigma^2}}$ **(3)**

where J(τ – τ') is the IRF, and σ is the standard deviation; $a$ is the bunching factor, while $\tau_1$ and $\tau_2$ are the antibunching and bunching time, respectively.

From the convoluted fitting function in (Eq. 1), an antibunching dip or $g^{(2)}_{meas}(0)$) of $0.16 \pm 0.01$, with the deduced antibunching time value of $\tau_1 \sim 3$ ns was derived (red solid line). The $g^{(2)}_{meas}(0)$ value is well below 0.5 and thus clearly indicates the single photon nature of the emitter.[28] Without the contribution of the system jitter, we achieved relatively similar result, with $g^{(2)}(0)$ of $0.15 \pm 0.01$ (blue dash line). This means that the small deviation from zero is mainly caused by the contribution of fluorescence background within our laser spot.[29] Polarization PL measurements showed that the transition dipole moment of the defect center is perpendicular to the optical axis (supporting information figure S1a). The emitter also exhibits high brightness with a saturated PL intensity in excess of $1 \times 10^6$ counts/sec (supporting information figure S1b).

Although off-resonant excitation is convenient due to its relative insensitivity to excitation wavelength, the excited electron is required to vibrationally relax before spontaneous emission takes place (figure 2a). Resonant excitation, on the other hand, enables coherent access to manipulation of quantum states, and is a practical means to realize photon indistinguishability. We therefore proceeded to investigate the emitter of interest resonantly. The Ti:Sap laser wavelength, frequency-stabilized by a high-resolution external cavity (WS6, HighFinesse), was scanned across the ZPL, and the fluorescence signal was collected by a single photon counting avalanche photodiode (SPCAPD). To minimize laser scattering, we chose to spectrally filter out the residual pump laser with a 830 nm longpass filter and collect the red-shifted photons from the PSB (highlighted in the figure 1c), instead of using the cross-polarization technique.[30] Figure 2b shows a resonant PLE plot obtained by averaging five consecutive scans over the same 14 GHz range with 70 MHz resolution (each scan lasting for ~ 2 minutes). The laser power was kept at 150 nW to prevent power-induced broadening. A single broad peak was observed at 766.186 nm, with a Gaussian FWHM of $6.3 \pm 0.3$ GHz. As discussed above, due to a permanent transition dipole moment of the emitter, spectral diffusion was expected to be observed, and such spectral fluctuations when averaged out, resulted in a broad Gaussian lineshape with a FWHM of 6.3 ± 0.3 GHz.[28] The average interval between two consecutive jumps appeared to be slow, on the order of milliseconds. Additional scans performed over a wider frequency range reveal numerous intermittent peaks with Gaussian linewidths of ~700 – 1200 MHz, randomly distributed in the scan window (figure 2c). The Gaussian fit peaks corresponded to the optical resonance of the emitters. The widths of these peaks are significantly smaller than the time-averaged value of ~ 6.3 GHz seen in figure 2b, which indicates that the emitter is stable but exhibits rapid spectral jumps during excitation. Spectral jumps are expected from emitters with permanent dipole moment such as the antisite vacancy defect $N_BV_N$, which has been suggested as the atomic structure of the hBN SPEs[5] studied in this work. We note, however, that some of the intermittent spectral features are not Gaussian, indicating a transition to a dark state (indicated by grey arrows in figure 2c). These measurements were recorded only with a single (resonant) laser. However, the emission instabilities seen in figure 2b-c may be mitigated, once the level structure of the emitters is fully understood, by a co-incident laser used to repump the defects, as has been done previously for the NV centers in diamond.[31]

To compare the measured peak widths to the Fourier-transform (FT) lifetime limited linewidth, we conducted time-resolved PL measurements of the excited state lifetime using a pulsed excitation source. A lifetime of 3.6 ± 0.1 ns was extracted by fitting a single exponential function to the experimental data shown in figure 2d, yielding a FT limited linewidth of ~ 44 MHz, which is over an order of magnitude narrower than the measure linewidths in figure 2b-c. The observed broadening and spectral jumps are attributed primarily to low quality of the hBN host material and the chemically exfoliated, drop-casted flakes used in the present study. Currently available hBN crystals are known to host impurities introduced during synthesis processes.[32, 33] Such impurities can undergo charge transitions during optical excitation, which can in turn give rise to intense local electric field fluctuations that interact strongly with the permanent transition dipole moment of emitters and cause spectral diffusion.[34] The latter can, in principle, be suppressed using dynamic Stark shift feedback techniques that have been demonstrated previously using nitrogen vacancy (NV) centers in diamond.[35] A more direct approach is to prevent the underlying problem by improving the crystal purity of the hBN material. Indeed, initial PLE measurements from NV centers in nanodiamonds showed GHz linewidths,[36, 37] and subsequent realization of high quality single crystal material yielded improved, stable and FT limited lines.

To characterize the emission dynamics further, we recorded the PL intensity as a function of time using a fixed excitation laser tuned to the emission resonance of 766.186 nm (red trace in figure 3a). Emission intermittency (blinking) is clearly evident when the laser is resonant with the optical transition of the emitter, and gives rise to photon bursts[38] in the time-resolved PL signal. For reference, off-resonance excitation with the laser wavelength detuned by 2 nm is shown as a black curve. In this case, the emitter was not excited and only a steady state, low intensity background is observed.

The time-resolved PL trace enables quantification of the spectral diffusion time. By setting a threshold value of 1800 counts/sec (grey dashed line) to separate the ON and OFF-resonance times ($\tau_{on}$ and $\tau_{off}$, respectively), the $\tau_{on}$ / $\tau_{off}$ ratio is equal to 0.47. This value indicates that the average amount of time during which the emitter is detuned from the excitation field is about twice as long as the ON-resonance time. By binning the ON-resonance time intervals (figure 3b), we obtain an average spectral diffusion time, $\tau_{avg\ spec\ diff}$, of approximately $102 \pm 65$ ms. We note, however, that spectral diffusion at the millisecond time scale cannot explain the line broadening observed in figure 2. Recent results show that emitters in hBN exhibit ultrafast spectral diffusion at a time scale of a few μs, and a coherence time of ~ 81 ps.[17] Further detailed studies are needed to fully understand and circumvent spectral diffusion of emitters in hBN.

With an average spectral diffusion time of ~ 100 ms, we demonstrated that an ON-resonance confocal PL map can be acquired, showing a clear bright spot in the center of the map (figure 3c, left panel), and the absence of background emissions present in the off-resonance confocal map shown in figure 1b. In contrast, when the resonant excitation laser was detuned by 2 nm, no PL signal is observed (figure 3c, right panel).

Finally, to confirm the on-resonance single photon purity, we acquired an on-resonance antibunching curve from the emitter using a laser power of 1 μW (figure 3d). An acquisition time of three hours was needed to achieve an adequate signal-to-noise ratio, due to the spectral diffusion and blinking which also prevented the observation of Rabi oscillations in our data. A fit based on the Gaussian-convoluted three-level model (red solid line) resulted in a $g^{(2)}(0)$ value of $0.11 \pm 0.01$, which is comparable to that from off-resonance excitation (figure 1d), and confirms the single photon nature of the emitter. The non-zero dip was due to a significant contribution of timing-jitter. By excluding the contribution of the jitter, however, the antibunching dip of $\sim 0 \pm 0.01$ is deduced, suggesting high single-photon purity of the source. Notably, the on-resonance antibunching time (~$0.87 \pm 0.02$ ns) is less than a third of the off-resonance antibunching time ($3.0 \pm 0.1$ ns).

To summarize, we demonstrated resonant excitation of a quantum emitter in hBN. The emitter has a time-averaged optical linewidth of ~ 6 GHz. Using individual scans, we were able to resolve narrower transitions, down to ~ 700 MHz, despite the presence of spectral diffusion. An average spectral diffusion time of ~ 100 ms was observed, which is sufficiently long to realize more complex experiments such as the Hong-Ou-Mandel two photon quantum interference effect.[39] Measurement of the second order autocorrelation function under resonant excitation was demonstrated, showing a single-photon purity of the source. Our results provide important insights into coherent properties of quantum emitters in hBN, and motivate further spectroscopic and materials engineering works aimed at improving the optical linewidths of quantum emitters in hBN.

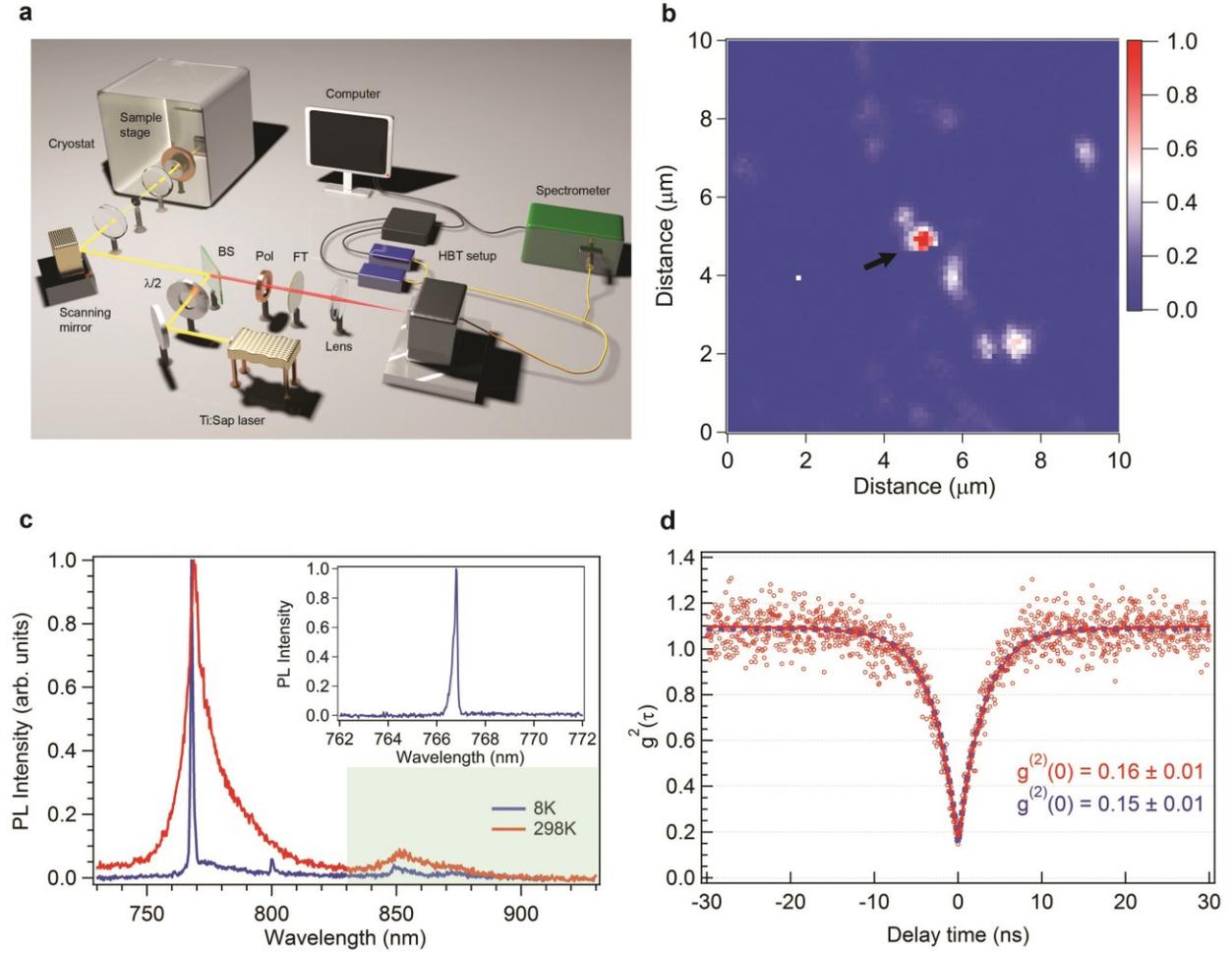

**Figure 1. a)** Cryogenic confocal PL setup. HBT: Hanbury-Brown and Twiss; BS: beamsplitter; FT: band-pass or long-pass filters; λ/2: half-wave plate; Pol: linear polarizer. **b)** Confocal PL map recorded with 700-nm laser excitation at 300 μW. The bright spot corresponds to a single emitter. The measurement was acquired at 8 K. **c)** Normalized PL spectrum taken from the same emitter at 8 K (blue trace) and 298 K (red trace) with a 300g/mm grating. The green-highlighted box indicates the collected spectral range for the PLE experiment in figure 2. The inset shows a higher resolution spectrum taken from the same emitter (with a 1800g/mm grating). **d)** Second-order autocorrelation function (red open circles) acquired for the emitter using a 700 nm laser at 100 μW power as the excitation source, acquired for 5 minutes. The measurement was conducted at 8 K. The red solid line is the fitting for the $g^{(2)}(0)$ function using a three-level model convoluted by a Gaussian jitter response (see main text). The $g^{(2)}(0)$ value of 0.16 ± 0.01, without any background correction, indicates that the emission is from a single emitter. The blue dash line shows the response of the three-level system only, yielding $g^{(2)}(0)$ value of 0.15 ± 0.01. A band-pass filter was used in the measurements of confocal PL in (b) and the photon second order autocorrelation function (d) to minimize the background PL contribution.

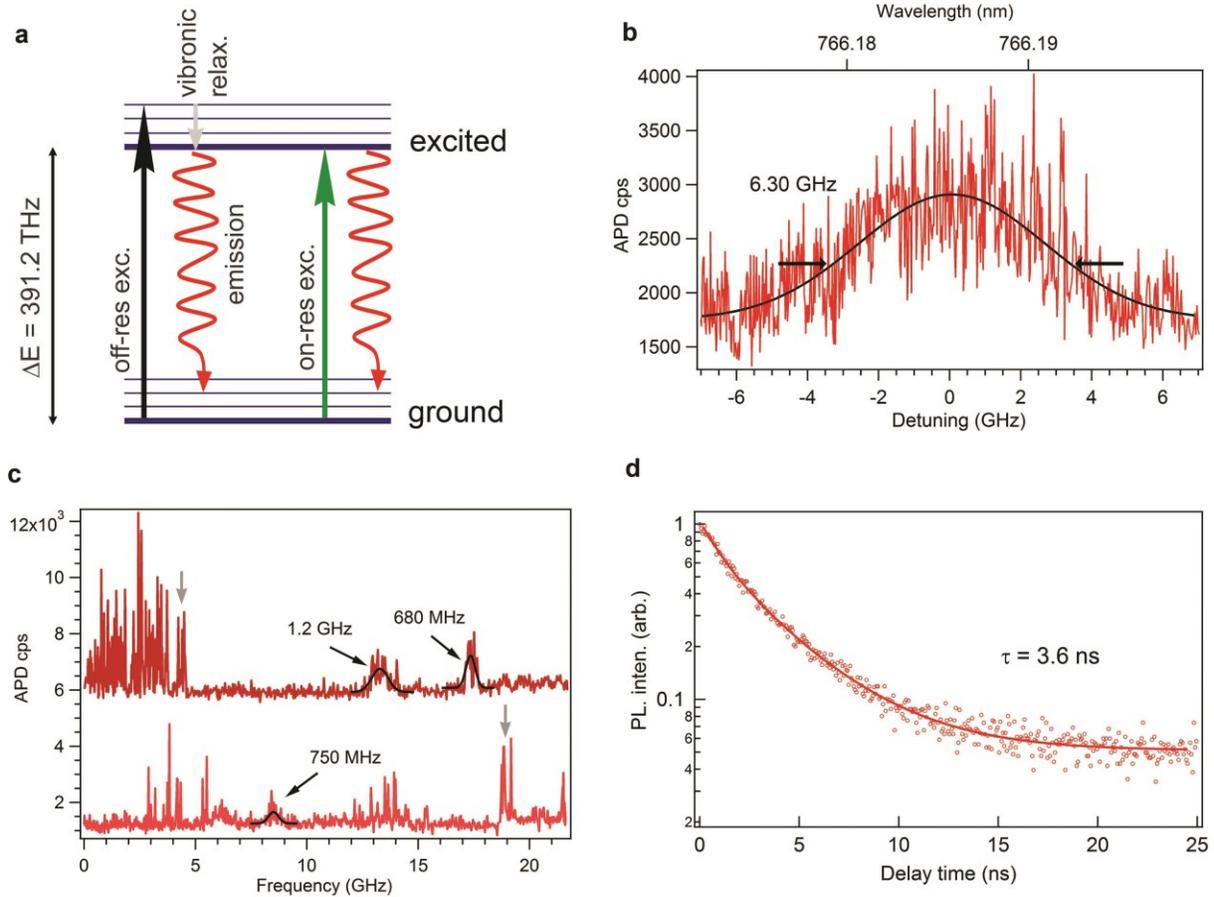

**Figure 2. a)** Simplified diagram of the hBN emitter where the excited state can be accessed via either off-resonance or on-resonance excitation, with the former pathway on the left, and the latter pathway on the right. Black and grey arrows, indicate excitation towards the higher vibronic states, followed by vibronic relaxation towards the excited ground state. The green arrow indicates on-resonance excitation, followed by spontaneous emission denoted by the wavy red arrow in both pathways. **b)** Resonance photoluminescence excitation measurements on the single emitter with a ZPL peak at 766.186 nm. The excitation power used was 150 nW. Only photons from the PSB were collected using a long pass filter. The experimental data is plotted as the red trace. Five repetitive scans were averaged out to get the final data. The data was fit with either a Gaussian function (black solid line). The measurement was done at 8 K. **c)** Additional survey PLE scans showing multiple local maxima with FWHM below 2 GHz. The grey arrows show representative spectral features that are not Gaussian, indicating a transition to a dark state. **d)** Time-resolved PL measurements (red open circles) of the same single emitter measured at room temperature. A single-exponential fit gives rise to a lifetime of 3.6 ns for the emitter's excited state. The measurement was done with a 675 nm pulsed laser (100 μW, 10 MHz repetition rate, 100 ps pulse width).

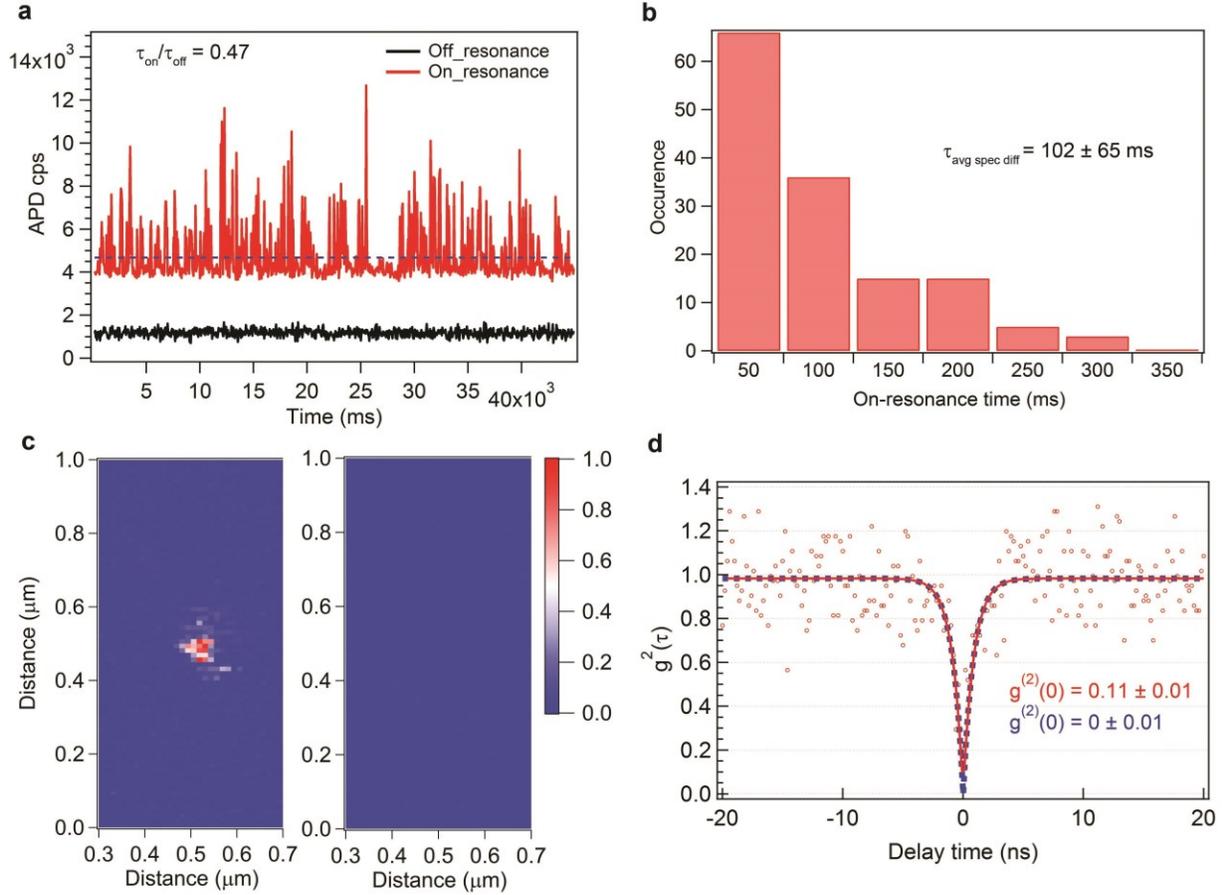

**Figure 3. (a)** PL intensity vs time for on-resonance (red line) and 2 nm detuned (black) excitation of the same emitter. The blue dash line represents the cut-off threshold (1800 count/sec) for calculating $\tau_{on}$ and $\tau_{off}$. The $\tau_{on} / \tau_{off}$ ratio was calculated to be 0.47. The data are vertically shifted for clarity. **(b)** Histogram of on-resonance time extracted from on-resonance trace of (a). The calculated average spectral diffusion time, $\tau_{avg\ spec\ diff} = 102 \pm 65$ ms. **(c)** Confocal PL map with the laser staying on-resonance (left panel) and 2 nm detuned (right panel) from the resonance. The measurements in (a) and (b) were carried out at the excitation power of 150 nW. **d)** On-resonance photon second-order correlation function (red open circles) acquired for the emitter at 1 µW excitation power for three hours. The red solid line is the fitting for the $g^{(2)}_{meas}(0)$ function using the Gaussian-convoluted three-level model, resulting in an antibunching dip value of $0.11 \pm 0.01$. The blue dash line concerns the response of the three-level system only, suggesting an antibunching dip value of $0 \pm 0.01$. All the measurement was conducted at 8 K.

AUTHOR INFORMATION

**Corresponding Author**

*Email: trongtoan.tran@student.uts.edu.au

*Email: igor.aharonovich@uts.edu.au

*Email: milos.toth@uts.edu.au

**Author Contributions**

The manuscript was written through contributions of all authors.

**Notes**

The authors declare no competing financial interest.

ACKNOWLEDGMENT

Financial support from the Australian Research Council (DE130100592), FEI Company, the Asian Office of Aerospace Research and Development grant FA2386-15-1-4044 are gratefully acknowledged. This research is supported in part by an Australian Government Research Training Program (RTP) Scholarship.

REFERENCES

1. Aharonovich, I.; Englund, D.; Toth, M. Solid-State Single-Photon Emitters. *Nat. Photon.* 2016, 10, 631-641.
2. Akselrod, G. M.; Argyropoulos, C.; Hoang, T. B.; Ciracì, C.; Fang, C.; Huang, J.; Smith, D. R.; Mikkelsen, M. H. Probing the Mechanisms of Large Purcell Enhancement in Plasmonic Nanoantennas. *Nat. Photon.* 2014, 8, 835-840.
3. Khasminskaya, S.; Pyatkov, F.; Słowik, K.; Ferrari, S.; Kahl, O.; Kovalyuk, V.; Rath, P.; Vetter, A.; Hennrich, F.; Kappes, M. M.; Gol'tsmanG; KorneevA; Rockstuhl, C.; Krupke, R.; Pernice, W. H. P. Fully Integrated Quantum Photonic Circuit with an Electrically Driven Light Source. *Nat. Photon.* 2016, advance online publication.
4. Chu, X.-L.; Götzinger, S.; Sandoghdar, V. A Single Molecule as a High-Fidelity Photon Gun for Producing Intensity-Squeezed Light. *Nat. Photon.* 2017, 11, 58-62.

5. Tran, T. T.; Bray, K.; Ford, M. J.; Toth, M.; Aharonovich, I. Quantum Emission from Hexagonal Boron Nitride Monolayers. *Nat. Nanotechnol.* 2016, 11, 37-41.
6. Martínez, L. J.; Pelini, T.; Waselowski, V.; Maze, J. R.; Gil, B.; Cassabois, G.; Jacques, V. Efficient Single Photon Emission from a High-Purity Hexagonal Boron Nitride Crystal. *Phys. Rev. B* 2016, 94, 121405.
7. Grosso, G.; Moon, H.; Lienhard, B.; Ali, S.; Efetov, D. K.; Furchi, M. M.; Jarillo-Herrero, P.; Ford, M. J.; Aharonovich, I.; Englund, D. Tunable and High Purity Room-Temperature Single Photon Emission from Atomic Defects in Hexagonal Boron Nitride. *arXiv preprint arXiv:1611.03515* 2016.
8. Tran, T. T.; Elbadawi, C.; Totonjian, D.; Lobo, C. J.; Grosso, G.; Moon, H.; Englund, D. R.; Ford, M. J.; Aharonovich, I.; Toth, M. Robust Multicolor Single Photon Emission from Point Defects in Hexagonal Boron Nitride. *ACS Nano* 2016, 10, 7331-7338.
9. Li, X.; Shepard, G. D.; Cupo, A.; Camporeale, N.; Shayan, K.; Luo, Y.; Meunier, V.; Strauf, S. Nonmagnetic Quantum Emitters in Boron Nitride with Ultranarrow and Sideband-Free Emission Spectra. *ACS Nano* 2017, 11, 6652-6660.
10. Jungwirth, N. R.; Calderon, B.; Ji, Y.; Spencer, M. G.; Flatté, M. E.; Fuchs, G. D. Temperature Dependence of Wavelength Selectable Zero-Phonon Emission from Single Defects in Hexagonal Boron Nitride. *Nano Lett.* 2016, 16, 6052-6057.
11. Chejanovsky, N.; Rezai, M.; Paolucci, F.; Kim, Y.; Rendler, T.; Rouabeh, W.; Fávaro de Oliveira, F.; Herlinger, P.; Denisenko, A.; Yang, S.; Gerhardt, I.; Finkler, A.; Smet, J. H.; Wrachtrup, J. Structural Attributes and Photodynamics of Visible Spectrum Quantum Emitters in Hexagonal Boron Nitride. *Nano Lett.* 2016, 16, 7037-7045.
12. Exarhos, A. L.; Hopper, D. A.; Grote, R. R.; Alkauskas, A.; Bassett, L. C. Optical Signatures of Quantum Emitters in Suspended Hexagonal Boron Nitride. *ACS Nano* 2017, 11, 3328.
13. Kianinia, M.; Regan, B.; Tawfik, S. A.; Tran, T. T.; Ford, M. J.; Aharonovich, I.; Toth, M. Robust Solid-State Quantum System Operating at 800 K. *ACS Photonics* 2017, 4, 768-773.
14. Tran, T. T.; Wang, D.; Xu, Z.-Q.; Yang, A.; Toth, M.; Odom, T. W.; Aharonovich, I. Deterministic Coupling of Quantum Emitters in 2d Materials to Plasmonic Nanocavity Arrays. *Nano Lett.* 2017, 17, 2634-2639.
15. Schell, A. W.; Takashima, H.; Tran, T. T.; Aharonovich, I.; Takeuchi, S. Coupling Quantum Emitters in 2d Materials with Tapered Fibers. *ACS Photonics* 2017, 4, 761-767.
16. Jungwirth, N. R.; Fuchs, G. D. Optical Absorption and Emission Mechanisms of Single Defects in Hexagonal Boron Nitride. *Phys. Rev. Lett.* 2017, 119, 057401.
17. Sontheimer, B.; Braun, M.; Nikolay, N.; Sadzak, N.; Aharonovich, I.; Benson, O. Photodynamic of Quantum Emitters in Hexagonal Boron Nitride Revealed by Low Temperature Spectroscopy. *arXiv preprint arXiv:1704.06881* 2017.
18. Becker, J. N.; Gorlitz, J.; Arend, C.; Markham, M.; Becher, C. Ultrafast All-Optical Coherent Control of Single Silicon Vacancy Colour Centres in Diamond. *Nat. Commun.* 2016, 7, 13512.
19. Zhou, Y.; Rasmita, A.; Li, K.; Xiong, Q.; Aharonovich, I.; Gao, W.-b. Coherent Control of a Strongly Driven Silicon Vacancy Optical Transition in Diamond. *Nat. Commun.* 2017, 8, 14451.
20. Taminiau, T. H.; Stefani, F. D.; Segerink, F. B.; Van Hulst, N. F. Optical Antennas Direct Single-Molecule Emission. *Nat. Photon.* 2008, 2, 234-237.

21. Grandi, S.; Major, K. D.; Polisseni, C.; Boissier, S.; Clark, A. S.; Hinds, E. A. Quantum Dynamics of a Driven Two-Level Molecule with Variable Dephasing. *Phys. Rev. A* 2016, 94.
22. Faez, S.; Türschmann, P.; Haakh, H. R.; Götzinger, S.; Sandoghdar, V. Coherent Interaction of Light and Single Molecules in a Dielectric Nanoguide. *Phys. Rev. Lett.* 2014, 113, 213601.
23. Englund, D.; Majumdar, A.; Faraon, A.; Toishi, M.; Stoltz, N.; Petroff, P.; Vuckovic, J. Resonant Excitation of a Quantum Dot Strongly Coupled to a Photonic Crystal Nanocavity. *Phys. Rev. Lett.* 2010, 104, 073904.
24. Faraon, A.; Fushman, I.; Englund, D.; Stoltz, N.; Petroff, P.; Vuckovic, J. Coherent Generation of Non-Classical Light on a Chip Via Photon-Induced Tunnelling and Blockade. *Nat. Phys.* 2008, 4, 859-863.
25. Tran, T. T.; Zachreson, C.; Berhane, A. M.; Bray, K.; Sandstrom, R. G.; Li, L. H.; Taniguchi, T.; Watanabe, K.; Aharonovich, I.; Toth, M. Quantum Emission from Defects in Single-Crystalline Hexagonal Boron Nitride. *Phys. Rev. Appl.* 2016, 5, 034005.
26. Bourrellier, R.; Meuret, S.; Tararan, A.; Stéphan, O.; Kociak, M.; Tizei, L. H. G.; Zobelli, A. Bright Uv Single Photon Emission at Point Defects in H-Bn. *Nano Lett.* 2016, 16, 4317-4321.
27. Aharonovich, I.; Castelletto, S.; Simpson, D. A.; Greentree, A. D.; Prawer, S. Photophysics of Chromium-Related Diamond Single-Photon Emitters. *Phys. Rev. A* 2010, 81, 043813.
28. Rabeau, J. R.; Stacey, A.; Rabeau, A.; Prawer, S.; Jelezko, F.; Mirza, I.; Wrachtrup, J. Single Nitrogen Vacancy Centers in Chemical Vapor Deposited Diamond Nanocrystals. *Nano Lett.* 2007, 7, 3433-3437.
29. Neu, E.; Agio, M.; Becher, C. Photophysics of Single Silicon Vacancy Centers in Diamond: Implications for Single Photon Emission. *Opt. Express* 2012, 20, 19956-19971.
30. Sipahigil, A.; Jahnke, K. D.; Rogers, L. J.; Teraji, T.; Isoya, J.; Zibrov, A. S.; Jelezko, F.; Lukin, M. D. Indistinguishable Photons from Separated Silicon-Vacancy Centers in Diamond. *Phys. Rev. Lett.* 2014, 113, 113602.
31. Robledo, L.; Bernien, H.; van Weperen, I.; Hanson, R. Control and Coherence of the Optical Transition of Single Nitrogen Vacancy Centers in Diamond. *Phys. Rev. Lett.* 2010, 105.
32. Yin, J.; Li, J.; Hang, Y.; Yu, J.; Tai, G.; Li, X.; Zhang, Z.; Guo, W. Boron Nitride Nanostructures: Fabrication, Functionalization and Applications. *Small* 2016, 12, 2942-2968.
33. Li, L. H.; Chen, Y. Atomically Thin Boron Nitride: Unique Properties and Applications. *Adv. Funct. Mater.* 2016, 26, 2594-2608.
34. Wolters, J.; Sadzak, N.; Schell, A. W.; Schröder, T.; Benson, O. Measurement of the Ultrafast Spectral Diffusion of the Optical Transition of Nitrogen Vacancy Centers in Nano-Size Diamond Using Correlation Interferometry. *Phys. Rev. Lett.* 2013, 110.
35. Acosta, V. M.; Santori, C.; Faraon, A.; Huang, Z.; Fu, K. M. C.; Stacey, A.; Simpson, D. A.; Ganesan, K.; Tomljenovic-Hanic, S.; Greentree, A. D.; Prawer, S.; Beausoleil, R. G. Dynamic Stabilization of the Optical Resonances of Single Nitrogen-Vacancy Centers in Diamond. *Phys. Rev. Lett.* 2012, 108.
36. Zhao, H. Q.; Fujiwara, M.; Okano, M.; Takeuchi, S. Observation of 1.2-Ghz Linewidth of Zero-Phonon-Line in Photoluminescence Spectra of Nitrogen Vacancy Centers in Nanodiamonds Using a Fabry-Perot Interferometer. *Opt. Express* 2013, 21, 29679-29686.
37. Tamarat, P.; Gaebel, T.; Rabeau, J. R.; Khan, M.; Greentree, A. D.; Wilson, H.; Hollenberg, L. C. L.; Prawer, S.; Hemmer, P.; Jelezko, F.; Wrachtrup, J. Stark Shift Control of Single Optical Centers in Diamond. *Phys. Rev. Lett.* 2006, 97, 083002.

38. Kumar, S.; Brotóns-Gisbert, M.; Al-Khuzheyri, R.; Branny, A.; Ballesteros-Garcia, G.; Sánchez-Royo, J. F.; Gerardot, B. D. Resonant Laser Spectroscopy of Localized Excitons in Monolayer Wse2. *Optica* 2016, 3, 882-886.
39. Hong, C. K.; Ou, Z. Y.; Mandel, L. Measurement of Subpicosecond Time Intervals between Two Photons by Interference. *Phys. Rev. Lett.* 1987, 59, 2044-2046.

Supporting information for

# Resonant excitation of quantum emitters from hexagonal boron nitride

*Toan Trong Tran,[1,*] Mehran Kianinia,[1] Minh Nguyen,[1] Sejeong Kim,[1] Zai-Quan Xu,[1] Alexander Kubanek,[2] Milos Toth,[1,*] and Igor Aharonovich[1,*]*

[1] School of Mathematical and Physical Sciences, University of Technology Sydney, Ultimo, NSW, 2007, Australia

[2] Institute for Quantum Optics, Ulm University, Albert-Einstein-Allee 11, D-89081 Ulm, Germany

[3] Center for Integrated Quantum Science and Technology (IQst), Ulm University, Albert-Einstein-Allee 11, D-89081 Ulm, Germany

In this supporting information, we present polarization spectra (figure S1a), and PL saturation (figure S1b).

To check the polarization of the light emitted from the center, we inserted a rotatable linear polarizer in the collection path. Figure S1a shows PL plots where maximum (red trace) and minimum (blue trace) PL intensity were achieved. The visibility of the emitter – an indication of how polarized the emission is – can be obtained with following expression:

$$VIS = \frac{I_{max} - I_{min}}{I_{max} + I_{min}} \quad (1)$$

where $I_{max}$ and $I_{min}$ are the maximum and minimum integrated emission intensity. A visibility of unity was determined which indicates that the emission is associated to a single transitional dipole moment, perpendicular to the optical axis.

The saturation curve for the emitter is shown in figure S1b. A fitting based on the below equation could be used to extract characteristic information about the emitter:

$$I = I_{max} \times P/(P + P_{sat}) \quad (2)$$

where $I_{max}$ and $P_{sat}$ are the maximum emission rate and excitation power at which saturation is reached, respectively. The fit produces values for $I_{max}$ and $P_{sat}$ of 1.3 Mcounts/sec and 1.6 mW, respectively.

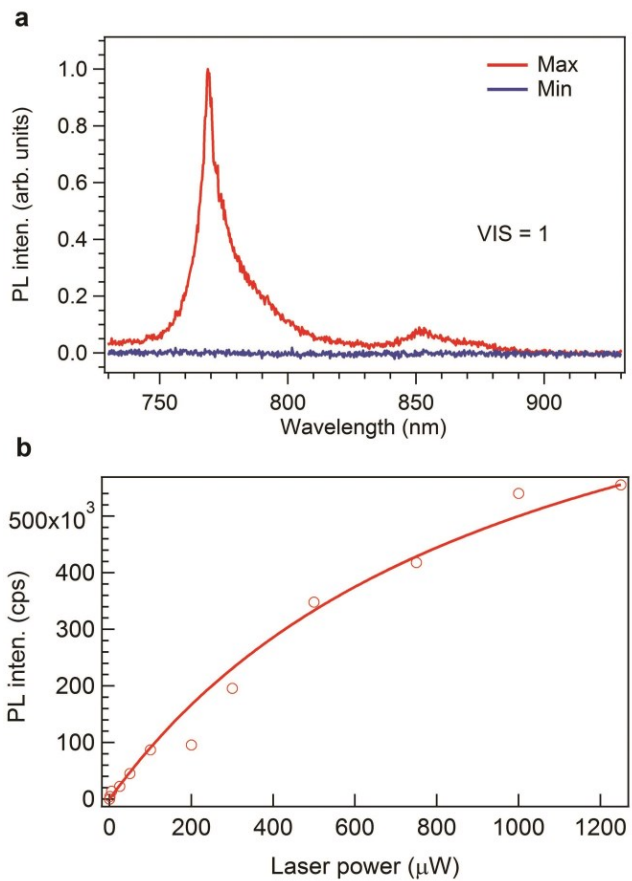

**Figure S1. a)** Spectra showing maximum (red trace) and minimum (blue trace) emission polarization from the emitter and taken with the use of a linear polarizer. The data was taken using excitation laser power of 300 µW with 5 s acquisition time. The visibility was determined to be at unity. **b)** Power-dependent fluorescence saturation curve (red open circles). The fit (solid red line) produces values of $I_{max}$ and $P_{sat}$ of 1.3 Mcounts/sec and 1.6 mW, respectively. The measurement was acquired with a band pass filter (760 ± 12) nm. All the measurements were conducted at room temperature.